\documentstyle[aps,12pt,epsfig]{revtex}

\def\be{\begin{equation}}
\def\ee{\end{equation}}
\def\disp{\displaystyle}
\def\foot{\footnotesize}

\begin{document}

\title{Anchoring of polymers by traps randomly placed on a line}
\bigskip

\author{S. Nechaev$^{1\dag}$, G. Oshanin$^2$ and A. Blumen$^3$}
\bigskip

\address{$^1$ UMR 8626, CNRS-Universit\'e Paris XI, LPTMS, Bat.100,
Universit\'e Paris Sud, \\ 91405 Orsay Cedex, France}
\address{$^2$ Physique Th\'eorique des Liquides,
Universit\'e Paris VI, T.16, 4 place Jussieu, \\ 75252 Paris Cedex 05, France}
\address{$^3$ Theoretische Polymerphysik, Universit\"at Freiburg,
Herman-Herder-Str. 3, \\ 79104 Freiburg, Germany}
\maketitle
\bigskip

\begin{abstract}
We study dynamics of a Rouse polymer chain, which  diffuses in a
three--dimensional space under the constraint that one of its ends,
referred to as the slip--link, may move only along a one--dimensional line
containing randomly placed, immobile, perfect traps.  For such a model we
compute exactly the time evolution of the probability $P_{\rm sl}(t)$ that
the chain slip--link will not encounter any of the traps until time $t$ and
consequently, that until this time the chain will remain mobile.
\end{abstract}
\vspace{0.5in}

\vspace{0.5in}

\noindent {\bf PACS:}
\vspace{0.5in}

\noindent Submitted to: {\sl Journal of Statistical Physics}
\vspace{2.5in}

\hrule
\bigskip

\noindent $^{\dag}$ also at L.D. Landau Institute for Theoretical Physics,
117940, Moscow, Russia.

\newpage

\section{Introduction}

The time evolution of the survival probability $P(t)$ of particles diffusing
in a $d$--dimensional space in the presence of immobile, randomly placed 
traps has been widely discussed in the physical and mathematical literature 
within the last two decades. An interest in  this problem has been inspired 
by the evident physical significance of the subject (excitation and charge 
motion, photoconductivity, photosynthesis). Further on, such an interest 
has been stimulated by an important observation \cite{bal} that $P(t)$ 
exhibits a non-mean-field long-time behavior, which is intimately related 
to the so-called Lifschitz singularities near the edge of the band in the 
density of states of a particle in quantum Lorentz gas and is reflected in 
the moment generating function of the so-called Wiener sausage 
\cite{don,pastur}. Later works (see, e.g., \cite{burl,3}) have also pointed 
out the relevance of the issue to percolation, self--avoiding random walks 
or self--attracting polymers, as well as to the anomalous behavior of the 
ground--state energy of a Witten's toy Hamiltonian in supersymmetric 
quantum mechanics \cite{sosiska}.

Various analytical techniques have been elaborated to calculate $P(t)$,
including an extension of the "optimal fluctuation" method \cite{bal},
different methods of evaluating upper and lower bounds (see, e.g.,
\cite{don,pastur,gp,kh,weiss}), Green functions approaches \cite{burl},
field--theoretic treatments \cite{3}, as well as a variety of 
mean--field--type descriptions (see \cite{fix,deutch,4} and references 
therein). These studies have revealed a two-stage decay pattern of the form
\be \label{p}
\ln P(t) \; \propto \; \left\{\begin{array}{lll}
- n_{\rm tr} \phi_d(t), & t_{\rm m}\ll t \ll t_{\rm c}, & \quad (A) \\
- n_{\rm tr}^{2/(d+2)} t^{d/(d+2)}, & t \gg t_{\rm c}, & \quad (B)
\end{array}\right.
\ee
where $n_{\rm tr}$ denotes the mean density of traps, $t_{\rm m}$ is a
microscopic time scale and  $t_{\rm c}$ denotes the crossover time between the
intermediate-- (A) and long--time (B) kinetic stages. Further on, the
function $\phi_{d}(t)$ appearing in Eq.(\ref{p}.A) defines the mean volume of
the so-called Wiener sausage (see, e.g. \cite{don})---i.e. the mean volume
swept by a diffusive spherical particle during time $t$. Its discrete-space
counterpart, i.e. an analog of $\phi_{d}(t)$  defined for lattice random walks,
is referred to as the mean number of distinct sites visited by a particle up
to the time $t$ (see \cite{blum3} for more details). The functional form of
$\phi_{d}(t)$ is different for different spatial dimensions $d$  and obeys:
\be \label{ph}
\phi_{d}(t) \; \propto \; \left\{\begin{array}{ll}
t^{1/2}, & d = 1\\ t/\ln(t), & d = 2\\
t, & d \geq 3 \end{array}\right.
\ee

The physical behavior underlying the kinetic regimes described by
Eqs.(\ref{p}.A) and (\ref{p}.B) has also been elucidated. It has been
understood that Eqs.(\ref{p}.A) and (\ref{p}.B)  are supported by completely
different realizations of random walk trajectories: The intermediate--time
behavior described by Eq.(\ref{p}.A) is associated with typical realizations
of random walk trajectories and is consistent with the predictions of the
mean--field, Smoluchowski--type approaches
\cite{burl,fix,deutch,4,blum1,blum2,blum3}; namely, $\phi_{d}(t) = \int^t dt'
k_{\rm Smol}(t')$ \cite{burl}, where $k_{\rm Smol}(t)$ is the so-called
Smoluchowski constant, which equals the diffusive current through the surface
of an immobile $d$-dimensional sphere. On the other hand, the long--time
asymptotical form in Eq.(\ref{p}.B) showing a slower time--dependence 
compared to the intermediate--time decay law, stems from the interplay 
between fluctuations in the spatial distribution of traps (namely, on the 
existence of rare but sufficiently large trap--free cavities), and atypical 
realizations of random walk trajectories which do not leave such cavities 
during the time of observation. Note also that  Eq.(\ref{p}.B) describes 
the anomalous long--time tail of the moment generating function for the 
Wiener sausage volume \cite{don}.

Due to the general interest in fractal structures as useful approximate models
of disordered media, and/or  anomalous diffusion, Eq.(\ref{p}) has been
extended \cite{4,blum1,blum2,blum3} to describe trapping of random walkers on
random or regular structures characterized, in the  general case, by a
non-integer spatial dimension $d_f$ and anomalous diffusion exponent $d_w$,
the latter being defined through the relation describing the time-dependence of
the second moment of the  particle's displacement, $\overline{r^2(t)} \sim
t^{2/d_w}$, where $d_{\omega}$ may be different from the value $d_{\omega}=2$,
which holds for conventional diffusive motion in Euclidean $d$--dimensional
space \cite{blum1,blum2,blum3}. For such systems heuristic agruments
\cite{4,dg,blum1,blum2,blum3,osh} suggest that $P(t)$ follows an asymptotic
behavior of the form
\be \label{pu}
\ln P(t) \; \propto \;
\left\{\begin{array}{lll}
-n_{\rm tr} \tilde{\phi}_{d}(t), & t_{\rm m}\ll t \ll t_{\rm c}, & \quad (A) \\
-n_{\rm tr}^{d_{w}/(d_{f}+d_{\omega})} \; t^{d_{f}/(d_{f}+d_{\omega})}, & t
\gg t_{\rm c}, & \quad (B)
\end{array}\right.
\ee
with
\be \label{phtilde}
\tilde{\phi}_{d}(t) \; \propto \;
\left\{\begin{array}{ll} t^{d_{f}/d_{\omega}}, & d_{f} < d_{\omega} \\
t/\ln(t), & d_{f} = d_{\omega} \\
t, & d_{f} > d_{\omega}
\end{array}\right.
\ee
Note, that here $d_f$ can attain integer values and  $d_{\omega}$ may be set
equal to $2$, which leads to conventional diffusive motion in Euclidean space;
then Eq.(\ref{pu}) reduces to Eq.(\ref{p}).

The decay patterns as in Eqs.(\ref{pu}) and (\ref{phtilde}) have been verified
numerically for different types of fractal systems, such as, e.g., Sierpinski
gaskets or percolation clusters \cite{blum1,blum2,blum3}. However, rigorous
results describing the evolution of $P(t)$ in systems showing anomalous
diffusion are lacking at present.

\begin{figure}
\centerline{\epsfig{file=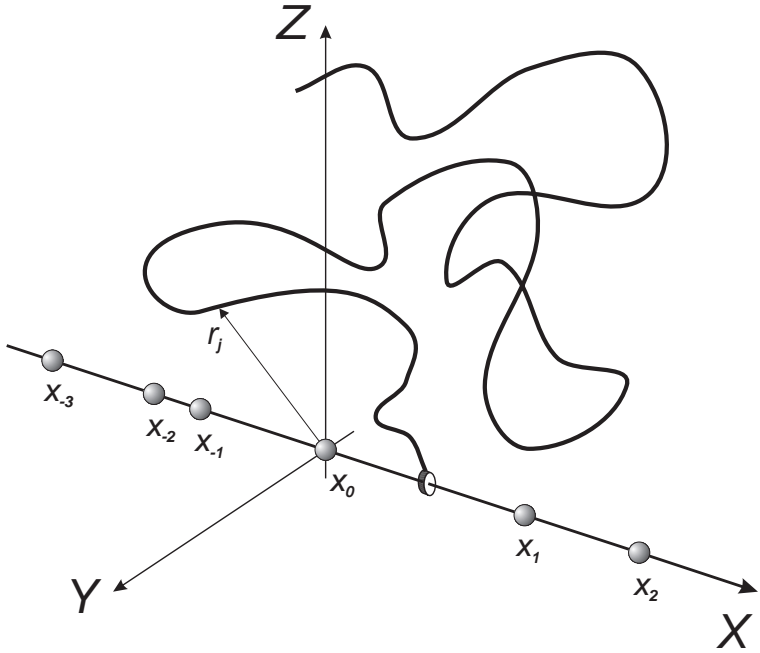,width=10cm}}
\caption{Polymer chain with an active particle attached to one of its
extremities (the slip--link).}
\label{fig:1}
\end{figure}

In this paper we discuss a particular case of trapping; we consider reactions
in systems in which the random motion of a particle moving in presence of
randomly placed, immobile traps is non--Markovian, but, nonetheless,
its survival probability can be determined exactly starting from first
principles. More specifically, we consider a situation in which a mobile
particle (referred to in what follows as a slip--link) is located at the 
end of a long polymer chain diffusing in three-dimensional space and is 
constrained to move along a one--dimensional line containing randomly 
placed perfect traps (\ref{fig:1}). The chain is modeled as a sequence of 
$N$ beads connected by phantom elastic springs and its dynamics is 
described within the framework of the customary Rouse model \cite{7}.

The paper is structured as follows: In Section 2 we formulate the model. In
Section 3 we discuss briefly the exact solution for the survival of a particle
which diffuses on a line in the presence of randomly distributed  traps, and
then rederive this solution in terms of a path--integral method. In Section 4
we describe, following \cite{5} the path integral formalism for evaluating the
measure of trajectories covered by a tagged bead of a Rouse chain. Next, we
show how such a formalism can be applied for the exact computation of the
probability that the slip--link is not trapped until time $t$. Finally, in
Section 5 we conclude with a summary and discussion of the  obtained results.

\section{The model}

Consider a polymer chain embedded in three--dimensional space and 
consisting of $N$ beads (Fig.\ref{fig:1}), which are connected sequentially 
by phantom harmonic springs of rigidity $K$. The rigidity can be also 
expressed as $K = 3 T/b^2$, where $T$ is the temperature (written in units 
of the Boltzmann constant $k_B$) and $b$ is the mean equilibrium distance 
between beads. Further more, the positions of all beads are denoted by 
$\vec{r}_j= (x_j,y_j,z_j)$, where the subscript $j$ enumerates the beads 
along the chain, $j=[0,N]$. All beads, except the slip--link ($j = 0$), may 
move freely in 3d. On contrary, we stipulate that the slip--link is 
constrained to move only along the $X$--axis, such that its position in 
space is given solely by the $X$--component of the vector $\vec{r}_0$, 
since $y_0$ and $z_0$ must always equal zero. Then the potential energy
$U(\{\vec{r}_j\})$ of such a chain is given by

\be \label{potential}
U\Big(\{\vec{r}_j\}\Big)=  \frac{K}{2}
\sum_{j=0}^{N-1} \Big(\vec{r}_{j+1}-\vec{ r}_j\Big)^2 = \frac{K}{2} \left[
\sum_{j=0}^{N-1} \Big(x_{j+1}-x_j\Big)^2 + \sum_{j=1}^{N-1} \Big\{
\Big(y_{j+1}-y_j\Big)^2 + \Big(z_{j+1}-z_j\Big)^2\Big\}\right]
\ee

Next, apart from the holonomic constraints imposed by the springs, the beads
experience the action of random forces $\vec{f}_{j}(t)$. These random forces
are assumed to be Gaussian and uncorrelated in time and space; their Cartesian
components $f_{j,\alpha}(t)$, where $\alpha = x,y,z$, obey:
\be \label{noise}
\begin{array}{l}
\overline{f_{j,\alpha}(t)} = 0, \\
\overline{f_{j,\alpha}(t) f_{j',\alpha'}(t')} = 2 \zeta T
\delta_{j,j'} \delta_{\alpha,\alpha'} \delta(t-t'),
\end{array}
\ee
where the bar stands for averaging over thermal histories and $\zeta$ is the
macroscopic friction coefficient.

In the absence of excluded--volume effects, the dynamics of the Rouse chain is
guided by the corresponding Langevin--Rouse equations \cite{7}
\be \label{RL}
\zeta \dot{\vec{ r}}_j =
-\frac{\delta U\Big(\{\vec{ r}_j\}\Big)}{\delta \vec{ r}_j}+
\vec{ f}_j(t),
\ee
where the dot denotes the time derivative. As one can verify readily, for the
potential energy given by Eq.(\ref{potential}) the equations for the
$\vec{r}_j$ decouple with respect to the Cartesian components. That is, the
dynamics of, say, $x_j$, is independent of $y_j$ and $z_j$ and reads
\be \label{rl}
\zeta \dot{x}_j = K (x_{j+1} + x_{j-1} - 2 x_{j}) +
f_{j,x}(t),
\ee
which equation holds for $j = 1,...,N-1$. On the other hand, the displacements
of the chain's extremities along the $X$-axis, i.e. $x_0$ and $x_{N}$, obey
\be
\zeta \dot{x}_0 = K (x_{1} - x_{0}) +
f_{0,x}(t),
\ee
and
\be
\zeta \dot{x}_N = K (x_{N-1} - x_{N}) +
f_{N,x}(t)
\ee
Hence, with regards to polymer dynamics, one faces an effectively
one--dimensional model.

Lastly, we suppose that the $N$--axis contains perfect, immobile traps, which
are placed at random positions with mean density $n_{\rm tr}$. The positions of
the traps are denoted by $\{X_n\}$, $-\infty<n<\infty$. According to our model,
the traps influence the dynamics of the polymer only by trapping (immobilising)
the slip--link at the encounter. The influence of the trap on the other beads (
i.e. such that $j\in[1,N]$) is indirect: as soon as the slip--link gets
immobilised by any of the traps, the chain becomes anchored as a whole due to
the links between the chain's monomers. Our aim is to compute exactly the time
evolution of the probability, $P_{\rm sl}(t)$, that a polymer chain with a
slip--link sliding along the $X$--axis will not encounter any of the traps 
(and thus will remain mobile until time $t$).

\section{Monomer trapping on a line}

It seems instructive to recall first the  time evolution of the survival
probability in the simplest case, for $N=0$, i.e. for a single chemically
active monomer diffusing on a one-dimensional line and reacting with randomly
placed, immobile traps. It is intuitively clear that for a Rouse chain
containing $N$ beads, we should recover (apart from the renormalization of
the diffusion coefficient) at sufficiently long times  the behavior predicted
for a single monomer, because for a finite chain the random motion of any of
the chain's beads ultimately follows the conventional diffusion of the chain's
center-of-mass \cite{7}. On the other hand, at shorter times substantial
deviations between the motion of the monomer and of the center--of--mass 
are to be found, due to the essentially non-diffusive characted of the 
slip--link motion, induced by the internal degrees of freedom of the 
polymer. This anomalous regime stemming from the internal relaxation modes 
of the chain will cause significant departures from the decay forms 
described by Eq.(\ref{p}).  The derivation of decay laws associated with 
this anomalous regime is the primary goal of the present paper and will be 
discussed in the next Section.

As one may expect, in one--dimensional systems the situation simplifies
considerably, since here the diffusive particle can react only with two
neighboring traps and thus cannot leave the intertrap interval. This renders
the problem exactly solvable, reducing it to the analysis of the particle
survival inside a finite interval, followed by averaging over the distribution
of the intertrap intervals. We outline such a calculation following the seminal
method of Ref.\cite{bal}.

\subsection{The one-dimensional exact solution.}

We calculate first the probability $\Psi(x,x(0),t|L)$ that a  diffusive
particle (whose diffusion coefficient is $D = T/\zeta$), which starts at
$x(0)$ will not encounter the traps at $X_0=0$ and at $X_1 = L$ until time $t$.
This probability follows as the solution of the following one--dimensional
boundary problem:
\be \label{1d}
\left\{
\begin{array}{l}
\disp \frac{\partial \Psi(x,x(0),t|L)}{\partial t}=
D \frac{\partial^2 \Psi(x,x(0),t|L)}{\partial x^2} \\
\disp \Psi(x,x(0),t|L)\big|_{x=0}=\Psi(x,x(0),t|L)\big|_{x=L}=0 \\
\Psi(x,x(0),t=0|L)=\delta(x - x(0))
\end{array}\right.
\ee
The solution of Eqs.(\ref{1d}) can be readily found by standard means and takes
the form of a Fourier series:
\be \label{series}
\Psi(x,x(0),t|L)=\disp \frac{2}{L} \; \sum_{n = 1}^{\infty}
\exp\left(-\frac{\pi^2 n^2 D t}{L^2}\right)
\; \sin\left(\frac{\pi n x}{L}\right)
\; \sin\left(\frac{\pi n x(0)}{L}\right),
\ee
Now, to compute the monomer survival probability we turn to the
position--averaged function
\be \label{psi}
\Psi(t|L)= \frac{1}{L} \; \int_0^L \int_0^L dx(0)\, dx\,
\Psi(x,x(0),t|L),
\ee
which can be  computed  from Eq.(\ref{series}) and reads
\be \label{ipsi}
\Psi(t|L) =
\disp \frac{8}{\pi^2} \; \sum_{l = 0}^{\infty} (2 l + 1)^{-2} \;
\exp\left(-\frac{\pi^2 (2 l + 1)^2 D t}{L^2}\right)
\ee
Next, the desired survival probability of the diffusive monomer $P(t)$ is
determined as the convolution
\be \label{surv}
P(t)=\int_0^{\infty} dL\, \Psi_{\rm mon}(t|L)\, {\cal P}(L)
\ee
where ${\cal P}(L)$ is the probability density of having a trap-free void of
length $L$. For a completely random (Poisson) placement of traps
${\cal P}(L)$ reads:
\be \label{poiss}
{\cal P}(L)= n_{\rm tr} \; \exp( - n_{\rm tr} L),
\ee
Consequently, one finds the following general expression determining the
monomer survival probability
\be \label{survival}
P(t)= \frac{8 n_{\rm tr}}{\pi^2} \; \sum_{l = 0}^{\infty}
(2 l+1)^{-2}\;\int_0^{\infty} dL\;
\exp\left(-\frac{\pi^2 (2 l+1)^2 D t}{L^2}-n_{\rm tr} L \right)
\ee
The asymptotical behavior of the expression in Eq.(\ref{survival}) has been
discussed in detail in Ref.\cite{weiss}; it has been shown that $P(t)$
follows a two--stage decay pattern as in Eq.(1). Explicitly on has
\be \label{inte}
P(t) \; \approx \;
\left\{\begin{array}{lll}
\exp\left(-4 n_{\rm tr} (D
t/\pi)^{1/2}\right), & t_{m} \ll t \ll t_{\rm c}, & \quad (A) \\
\exp\left(-3 (\pi^2 n_{\rm tr}^2 D t/4)^{1/3}\right), & t \gg t_{\rm c},
& \quad (B) \end{array} \right.
\ee
where the crossover time $t_{\rm c}$ separating two regimes obeys $t_{\rm c}
\approx 1/D n_{\rm tr}^2$, and consequently, can be large if $n_{\rm tr}$ is
small.

\subsection{Path--integral solution of the monomer trapping problem in 1D.}

In this subsection we rederive the solution of the monomer trapping problem
in terms of the path--integral formalism, which will be later used to 
determine the trapping kinetics of the slip--link. To do this, we will 
proceed as follows:

We first write the solution of Eq.(\ref{1d}) as an
integral over Brownian paths $x(\tau)$:
\be \label{1d_1}
\Psi\Big(x,x(0),t|L\Big)= \frac{1}{2\sqrt{\pi Dt}} \left.
\int\limits_{x(0)}^{x}{\cal  D}\left\{x(\tau)\right\}
\exp\left\{-\frac{1}{4 D}\int_0^t d\tau
\left(\frac{\partial x(\tau)}{\partial \tau}\right)^2\right\}
\right|_{0 < x(\tau)< L, \; \tau \in [0; t]},
\ee
where ${\cal D}\{x(\tau)\}$ denotes integration over the monomer
trajectories $x(\tau)$, the exponential is the standard Wiener measure,
while the subscript $0<x(\tau)<L, \; \tau \in [0,t]$ signifies that the
integral has to be calculated under the constraint that neither of the
monomer's trajectories leaves the interval $[0,L]$ within the time period
$[0,t]$.

Next, such a constraint can be automatically taken into account if we multiply
the integrand by the  step function depicted in the figure below
\bigskip

\unitlength=1.00mm
\special{em:linewidth 0.6pt}
\linethickness{0.6pt}
\begin{picture}(120.00,30.00)
\put(20.00,10.00){\line(0,1){10.00}}
\put(20.00,20.00){\line(1,0){20.00}}
\put(40.00,20.00){\line(0,-1){10.00}}
\put(60.00,10.00){\line(0,1){10.00}}
\put(60.00,20.00){\line(1,0){20.00}}
\put(80.00,20.00){\line(0,-1){10.00}}
\put(40.00,5.00){\makebox(0,0)[cc]{\small $(2m-1) L$}}
\put(60.00,5.00){\makebox(0,0)[cc]{\small $2m L$}}
\put(80.00,5.00){\makebox(0,0)[cc]{\small $(2m+1) L$}}
\put(60.00,10.00){\vector(0,1){20.00}}
\put(120.00,5.00){\makebox(0,0)[cc]{$x$}}
\put(62.00,30.00){\makebox(0,0)[lc]{$u(x)$}}
\put(57.00,20.00){\makebox(0,0)[cc]{\small $1$}}
\put(100.00,10.00){\line(0,1){10.00}}
\put(100.00,20.00){\line(1,0){10.00}}
\put(10.00,10.00){\vector(1,0){110.00}}
\put(115.00,20.00){\makebox(0,0)[cc]{$\cdots$}}
\end{picture}

Using the contour integral representation of such a step function \cite{8}
\be \label{step2}
u(x)=\frac{1}{2\pi i}\int\limits_{c-i\infty}^{c+i\infty}
\frac{d\lambda\; e^{-\lambda x}}{\lambda(1+e^{\lambda L})}=
\left\{\begin{array}{ll} 1 & \mbox{if $2m L< x <(2m+1)L$} \\
0 & \mbox{if $(2m+1)L< x <(2m+2) L$} \end{array}\right.
\ee
where $m=0,1,2, \ldots$,  we can rewrite Eq.(\ref{1d_1}) as
\be \label{1d_path}
\Psi\Big(x,x(0),t|L\Big)=\frac{1}{2\sqrt{\pi D t}}
\sum_{m=-\infty}^{\infty}\frac{1}{2\pi i}
\int\limits_{c-i\infty}^{c+i\infty}
\frac{d\lambda\; e^{\lambda x(0)}}{\lambda(1+e^{\lambda L})}
\int\limits_{x(0)}^{x} {\cal D}\left\{x(\tau)\right\}
\exp\Big[-S\{x(\tau)\}\Big],
\ee
where the action $S\{x(\tau)\}$ is given by
\be \label{1d_act}
S\{x(\tau)\}=\int_0^t d\tau\left\{\frac{1}{4 D}
\left(\frac{\partial x(\tau)}{\partial \tau}\right)^2+
\lambda \left(\frac{\partial x(\tau)}{\partial \tau}\right) \right\}
\ee

Now, to compute the path--integral in Eq.(\ref{1d_path}) with the quadratic
action in Eq.(\ref{1d_act}) we have  merely to define the action-minimizing
trajectory $\tilde{x}(\tau)$ and calculate the action corresponding to such a
trajectory. The action--minimizing trajectory is defined by the classical Euler
equation of motion, which for the action in Eq.(\ref{1d_act}) is simply
$$
\frac{d}{d \tau}\left(\frac{1}{2 D} \frac{d \tilde{x}(\tau)}{d\tau}+
\lambda\right)=0
$$
Integrating this equation subject to the conditions $\tilde{x}(\tau = 0) =
x(0)$ and $\tilde{x}(\tau = t) = x$, we find
\be \label{min1}
\tilde{x}(\tau)=(x - x(0)) \frac{\tau}{t} +x(0)
\ee
Consequently, the minimal action is given by
$$
S\{\tilde{x}(t)\} = \frac{(x - x(0))^2}{4 D t} +
\lambda (x - x(0))
$$
and hence, the formal solution of the boundary problem in Eq.(\ref{1d}) can be
written down as
\be \label{1d_sol}
\begin{array}{lll}
\disp \Psi(x,x(0),t|L) & = & \disp \frac{1}{2\sqrt{\pi D t}}
\sum_{m=-\infty}^{\infty} \frac{1}{2\pi i}
\int\limits_{c-i\infty}^{c+i\infty} \frac{d\lambda}{\lambda(1+e^{\lambda
L})} \exp\left\{-\frac{(x-x(0))^2}{4 D t}+\lambda x \right\} = \medskip \\
& = & \disp \frac{1}{2\sqrt{\pi D t}} \sum_{m=-\infty}^{\infty}
\exp\left\{-\frac{(x-x(0)-2m L)^2}{4 D t}\right\}
\end{array}
\ee
Next, using the well--known representation of Jacobi theta-function \cite{8}
$$
\sum _{m=-\infty}^{\infty}q^{(m+a)^2}=\theta_3\left(\pi a,\;
\exp\left(\frac{\pi^2}{\ln q}\right)\right)\;\ln^{-1/2}\frac{1}{q}
$$
and setting
\be \label{def}
q=\exp\left(-\frac{L^2}{ D t}\right); \qquad a=\frac{x-x(0)}{2L}
\ee
we may rewrite Eq.(\ref{1d_sol}) as
\be \label{const}
\begin{array}{lll}
\Psi(x,x(0),t|L)=\disp \frac{2}{L}
\sum_{m=-\infty}^{\infty} \exp\left(-\frac{{\cal D}\pi^2
m^2 t}{L^2}\right) \cos\left(\frac{\pi m (x-x(0))}{L}\right) =
\medskip \\ \disp \frac{2}{ L}
\sum_{m=-\infty}^{\infty} \exp\left(-\frac{ D \pi^2
m^2 t}{L^2}\right) \left(\underbrace{\sin\left(\frac{\pi m x}{L}\right)
\sin\left(\frac{\pi m x(0)}{L}\right)}_{\foot \rm part\; A}+
\underbrace{\cos\left(\frac{\pi m x}{L}\right)
\cos\left(\frac{\pi m x(0)}{L}\right)}_{\foot \rm part\;B}\right)
\end{array}
\ee
Note now that the contribution "A" (a product of two "odd"
sine--functions---see Eq.(\ref{const})) vanishes at $x=0$ and $x=L$. It thus
mirrors the absorbing boundary conditions relevant to the trapping problem,
while the second contribution "B" (a product of tho "even" cosine--functions)
corresponds to the totally reflecting boundary conditions. Moreover, the
expression in Eq.(\ref{const}) in which only the odd contribution "A" is taken
into account coincides with the result presented in Eq.(\ref{series}). It
follows that the asymptotical behavior of the monomer survival probability can
be evaluated in terms of the path--integral formalism by picking then the odd
contribution. We would like to note also that the construction we have employed
is nothing but the well known "mirror principle", widely used in electrostatics.

We conclude this subsection with the following prescription:

\noindent {\bf Prescription (the "mirror principle").} {\it To find the
solution of the diffusion problem  with the Dirichlet boundary condition at
the ends of the segment $[0,L]$, we have merely to: a) obtain the Green's
function solution $\Psi(x,x(0),t)$ on the full line $\{x,x(0)\}\in ]-\infty,
\infty[$; b)  restrict $x-x(0)$ to the segment $[0,L]$, perform the
replacement $x-x(0)\to x-x(0)-2mL$ and take the odd contribution to the sum
$\sum_{m=-\infty}^{\infty} \Psi(x-x(0)-2mL,t)$.}

\section{Trapping of the slip--link of a polymer}

To calculate the evolution of $P_{\rm sl}(t)$ in the case of a slip--link
attached to a polymer chain, we will proceed essentially along the lines of
the previous section. First, we  present the derivation of $\Psi_{\rm
sl}\Big(x_0,x_0(0), t\Big)$---the probability distribution for the
displacements of the slip--link on an infinite line without traps \cite{5}.
Then, using the above formulated prescription (the "mirror principle"), we
determine $\Psi_{\rm sl}\Big(x_0,x_0(0), t|L\Big)$---the probability that
the slip--link of the chain, which is initially located at some point
$x_0(0)$ inside the interval $[0,L]$, will not leave this interval until
time $t$. Finally, the desired probability $P_{\rm sl}(t)$  will be
obtained from $\Psi_{\rm sl}\Big(x_0,x_0(0), t|L\Big)$ by averaging over
the Poisson distribution of the interval's lengths.

\subsection{The probability distribution of the slip--link displacement}

In this subsection we outline, following the analysis of Ref.\cite{5}, the
steps involved in the derivation of the probability distribution
$\Psi_{\rm sl}\Big(x_0,x_0(0),t\Big)$ for the displacements of the slip--link 
on an infinite line.

This probability  can be written as
\be \label{aver}
\Psi_{\rm sl}\Big(x_0,x_0(0),t\Big)=\int_{-\infty}^{\infty}...
\int_{-\infty}^{\infty} dx_1 ... dx_N \Psi(x_0,x_1..,x_N,t)
\ee
where $\{x_0,x_1,..,x_N\}\equiv \{x_1(t),..,x_N(t)\}$ here and henceforth
denote the coordinates of all chain segments at time $t$, while
$\Psi(x_0,x_1,..,x_N,t)$ is the joint distribution function of the
$N$ beads of the polymer chain. In other words, $\Psi(x_{0},x_{1},...,
x_{N},t)$ is the probability of having the $x$--coordinates of the $N$
beads of the chain at the positions $\{x_{j}(t)\}$, provided that initially
they where at $\{x_{j}(0)\}$.

For the Rouse chain whose potential energy obeys Eq.(\ref{potential}), the
time evolution of the function $\Psi(x_{0},x_{1},..,x_{N},t)$ is governed by
the following Smoluchowski--Fokker--Planck equation \cite{7}:
\be \label{smol2}
\frac{\partial}{\partial t}\Psi(x_{0},x_1,..,x_N,t) =
D \sum_{j=0}^{N}\frac{\partial}{\partial x_j}
\left(\frac{\partial}{\partial x_j}+\beta\frac{\partial}{\partial x_j}
U(x_{0},x_1,..,x_N)\right)\Psi(x_{0},x_1,..,x_N,t)
\ee
where $\beta=1/T$ and $ D = T/\zeta$ is the diffusion constant of an individual
bead (monomer). Equation (\ref{smol2}) has to be solved subject to the initial
condition:
\be \label{in_bound1}
\disp \Psi(x_{0},x_1,..,x_N,t=0)=\prod_{j=0}^N \delta(x_j(t)-x_j(0))
\ee
Eqs.(\ref{smol2})--(\ref{in_bound1}) define completely the evolution function
$\Psi(x_{0},x_1,..,x_N,t)$.

The computation of the probability distribution $\Psi(x_{0},x_0(0),t)$ using
the path-integral formalism was first performed in \cite{5}. Let us recall the
main steps of this approach. First of all, it is expedient to cast
Eq.(\ref{smol2}) into the form of the $(N+1)$--dimensional Schr\"odinger--type
equation. This can be readily performed by making use the following ansatz:
\be \label{subst}
\Psi(x_{0},x_1,..,x_N,t)=\exp\left\{-\frac{K (N + 1)}{2 \zeta} t -
\frac{\beta}{2} U(x_{0},x_1,..,x_N)\right\}\Phi(x_{0},x_1,..,x_N,t),
\ee
where the function $\Phi(x_{0},x_1,..,x_N,t)$ satisfies
\be \label{schr}
\frac{\partial}{\partial t}\Phi(x_{0},x_1,..,x_N,t)=
D \sum_{j=0}^{N}\left(\frac{\partial^2}{\partial x_j^2}-
\beta^2 K^2 \left(x_{j-1}-2x_j+x_{j+1}\right)^2\right) \Phi(x_{0},x_1,..,x_N,t)
\ee
Now, by virtue of the Feynmann-Kac theorem, the formal solution of
Eq.(\ref{schr}) can be written down explicitly as the following
path--integral
\be \label{path}
\Phi(x_{0},x_1,..,x_N,t)=\frac{1}{{\cal N}}\int\limits_{x_0(0)}^{x_0}...
\int\limits_{x_N(0)}^{x_N} \prod\limits_{j = 0}^N
{\cal D}\left\{x_j(\tau)\right\}
\exp\Big[ - S\Big\{x_{0}(\tau),x_1(\tau),..,x_N(\tau)\}\Big]
\ee
where ${\cal N}$ is the normalisation constant and the action
$S\Big\{x_{0}(\tau),x_1(\tau),..,x_N(\tau)\Big\}$ has the form:
\be \label{action}
\begin{array}{l}
\disp S\Big\{x_{0}(\tau),x_1(\tau),..,x_N(\tau)\Big\} = \\
\disp \int\limits_0^t d\tau\left\{\frac{1}{4 D}\sum_{j=0}^{N}
\left(\frac{\partial x_j(\tau)} {\partial \tau}\right)^2 + D \beta^2 K^2
\sum_{j=1}^{N-1} \left(x_{j-1}(\tau)-2x_j(\tau) + x_{j+1}(\tau)\right)^2
\right\} \end{array}
\ee
In the subsequent calculations we will not specify the normalisation
constant and pre-exponential factors; all our results will hold thus only
to exponential accuracy.

Now,  we suppose following Ref.\cite{5} that the slip--link moves along some
prescribed trajectory $X_{\rm sl}(\tau)$; as a matter of fact, such a
constraint allows to symmetrize the boundary conditions for the
action--minimizing trajectory, which are otherwise different at different
chain's extremities.  Such a constraint can be taken into account by
multiplying the integrand in Eq.(\ref{path}) by a functional 
delta--function of the form (see Ref.\cite{5} for details) 
$$
\delta\Big(x_0(\tau)-X_{\rm sl}(\tau)\Big)=\int {\cal D}\{m(\tau)\}
\exp\left\{-i\int_0^t d\tau\, m(\tau)\Big(x_0(\tau)-
X_{\rm sl}(\tau)\Big)\right\},
$$
which means that the action in Eq.(\ref{action}) is replaced by an effective
action of the following form:
\be \label{action2}
\begin{array}{l}
\disp S'\Big\{x_{0}(\tau),x_1(\tau),...,x_N(\tau)\}
= \int_0^t d\tau\Bigg\{\frac{1}{4 D} \sum_{j=0}^{N}\left(\frac{\partial
x_j(\tau)} {\partial \tau}\right)^2 + \\
\disp D \beta^2 K^2 \sum_{j=1}^{N-1}
\left(x_{j-1}(\tau)-2x_j(\tau) + x_{j+1}(\tau)\right)^2 +
i m(\tau)\Big(x_0(\tau)-X_{\rm sl}(\tau)\Big)\Bigg\},
\end{array}
\ee
and one  has to perform afterwards an additional integration over the measure
${\cal D}\{m(\tau)\}$.

Now, the action in (\ref{path}) is minimal for classical trajectories
satisfying the Euler equation:
$$
\left\{\frac{d}{d\tau}\left(\frac{\partial}{\partial \dot{x}_j}\right)-
\frac{\partial}{\partial x_j}\right\} {\cal L}
\left(\dot{x}_0,\dot{x}_1,...,\dot{x}_N, x_0,x_1,..,x_N \right)=0;
\qquad j\in[0,N],
$$
where ${\cal L}$ is the Lagrangian function:
\be \label{Lag}
{\cal L} = \frac{1}{4 D} \sum_{j=0}^{N}\left(\frac{\partial x_j(\tau)}
{\partial \tau}\right)^2 + D \beta^2 K^2 \sum_{j=1}^{N-1}
\Big(x_{j-1}(\tau)-2x_j(\tau) + x_{j+1}(\tau)\Big)^2 +
i m(\tau)\Big((x_0(\tau)-X_{\rm sl}(\tau)\Big)
\ee
Turning to the continuous $j$--limit, one obtains the following Euler equation,
determining the optimal trajectories of the chain's monomers \cite{5}:
\be \label{min}
\left(\frac{\partial}{\partial\tau} - 2 K \beta D \frac{\partial^2}{\partial
j^2}\right) \left(\frac{\partial}{\partial\tau} +
2 K \beta D \frac{\partial^2}{\partial j^2}\right)
\tilde{x}_j(\tau)= 4 \, D \, i \, \delta(j) \, m(\tau),
\ee
which has to be solved subject to the boundary conditions
(see \cite{5})
\be \label{bound}
\left\{\begin{array}{l}
\disp \frac{\partial \tilde{x}_j(\tau)}{\partial j}\bigg|_{j=0,N}=0, \qquad
\frac{\partial^3 \tilde{x}_j(\tau)}{\partial j^3}\bigg|_{j=0,N}=0 \medskip \\
\disp \frac{\partial \tilde{x}_j(\tau)}{\partial \tau}=-2 K \beta D \;
\frac{\partial^2 \tilde{x}_j(\tau)}{\partial j^2}\bigg|_{\tau=0}, \qquad
\frac{\partial \tilde{x}_j(\tau)}{\partial \tau}= + 2 K \beta D \;
\frac{\partial^2 \tilde{x}_j(\tau)}{\partial j^2}\bigg|_{\tau=t}
\end{array}\right.
\ee

The action--minimizing trajectories $\tilde{x}_{j}(\tau)$, defined by the
boundary problem (\ref{min})--(\ref{bound}), are obtained explicitly \cite{5}
in form of a series expansion over the normal Rouse modes \cite{7}:
\be \label{tr}
\tilde{x}_j(\tau)=-\frac{2 i N }{\pi^2 K \beta}\sum_{p=1}^{\infty} p^{-2}
\cos\Big(\frac{\pi p j}{N}\Big)
\int_0^t d\tau'\, m(\tau')
\exp\left\{-\frac{|\tau-\tau'|}{\tau_{R}}  p^2\right\}
\ee
where $\tau_{R} = \zeta N^2/2 \pi^2 K$ is the largest fundamental relaxation
time of the harmonic chain, i.e., the so-called Rouse time. This time may be
interpreted as being the time needed for some local defect, e.g., kink, to
spread out diffusively  along the arclength of the chain.

Next, substituting the expression for the optimal trajectory in Eq.(\ref{tr})
into Eq.(\ref{Lag}) and performing the integration over
${\cal D}\{m(\tau)\}$, one arrives at the following general result \cite{5}:
\be \label{triu2}
\Psi_{\rm sl}(x_{0},x_{0}(0),t)=\frac{1}{{\cal N}}\int\limits_{x_{0}(0)}^{x_{0}}
{\cal D} \Big\{X_{\rm sl}(\tau)\Big\} \exp\left[ - \int_{0}^{t}
\left(\frac{dX_{\rm sl}(\tau)}{d\tau}\right)\, d\tau
\int_{0}^{t} \left(\frac{dX_{\rm sl}(\tau')}{d\tau'}\right)\,d\tau'
\; \phi(\tau - \tau')\right],
\ee
where $\phi(\tau - \tau')$ is given by
\be \label{i}
\phi(\tau - \tau') \; \approx \;
\left\{\begin{array}{lll}  \frac{N}{4 D}\, \delta(\tau - \tau'),  &
|\tau - \tau'| > \tau_{R}, & \quad (A) \\
\left(\frac{\beta K}{D}\right)^{1/2}\, |\tau - \tau'|^{-1/2}, &
|\tau - \tau'| < \tau_{R}, & \quad (B) \end{array} \right.
\ee
Equations (\ref{triu2}) and (\ref{i}) represent the desired generalization of
the classical Wiener result for the measure of Brownian particle trajectories
to the more complicated case of a particle attached to a diffusive Rouse chain.

Finally, in order to compute the probability $\Psi_{\rm sl}(x_0,x_0(0),t|L)$ that
the slip--link of the Rouse chain will remain until time $t$ within the interval
$[0,L]$, we make use of the "mirror principle" prescription of the previous
section. Multiplying the integrand in Eq.(\ref{triu2}) by a step--function, we
get
\be \label{1d_path_pol}
\Psi_{\rm sl}\Big(x_0,x_0(0),t|L\Big)=\frac{1}{{\cal N}}
\sum_{m=-\infty}^{\infty}\frac{1}{2\pi i}
\int\limits_{c-i\infty}^{c+i\infty}
\frac{d\lambda\; e^{\lambda x_0(0)}}{\lambda(1+e^{\lambda L})}
\int\limits_{x_{0}}^{x_{0}(0)} {\cal D}\Big\{X_{\rm sl}(\tau)\Big\}
\exp\left[-S\Big\{X_{\rm sl}(\tau)\Big\}\right]
\ee
where
\be \label{newaction}
S\Big\{X_{\rm sl}(\tau)\Big\} = \int_{0}^{t}
\left(\frac{dX_{\rm sl}(\tau)}{d\tau} \right)\, d\tau \int_{0}^{\tau}
\left(\frac{dX_{\rm sl}(\tau')}{d\tau'}\right)\, d\tau'\,\phi(\tau - \tau')+
\lambda \int_0^t d\tau \left(\frac{d X_{\rm sl}(\tau)}{d \tau}\right)
\ee

Below we discuss the asymptotical forms of $\Psi_{\rm sl}\Big(x_0,x_0(0),
t|L\Big)$ using Eqs.(\ref{1d_path_pol}) and (\ref{newaction}) in the limits
$|\tau-\tau'| < \tau_{R} $ and $|\tau-\tau'| > \tau_{R}$.

\subsection{Asymptotic behavior of $\Psi_{\rm sl}\Big(x_0,x_0(0),t|L\Big)$.}

We focus first on the behavior in the intermediate--time limit, $t < \tau_{R}$.
Note, however, that since $\tau_R \sim N^2$, for sufficiently long chains this
intermediate--time regime may last over quite an extended time interval.

It follows from Eqs.(\ref{1d_path_pol}) and (\ref{newaction}) that
for  $t < \tau_{R}$ the internal relaxations of the chain
are most important and lead to the following form of the action
\be \label{u}
S\Big\{X_{\rm sl}(\tau)\Big\}=\left(\frac{\beta K}{D}\right)^{1/2}
\int_0^t \left(\frac{\partial X_{\rm sl}(\tau)}{\partial \tau}\right)
d \tau \int_0^{\tau} \left(\frac{\partial X_{\rm sl}(\tau')}{\partial
\tau'}\right) \frac{d \tau'}{\sqrt{|\tau-\tau'|}}  + \lambda \int_0^t d\tau
\left(\frac{d X_{\rm sl}(\tau)}{d \tau}\right).
\ee
Now $S$ in Eq.\ref{u} is non-local and possesses non-Wiener scaling properties.
In particular, it yields for the mean--square displacement of any chain's bead,
including the slip--link, the following law \cite{dg,7,5}
\be \label{disp}
\overline{x_{j}^2(t)} \sim t^{1/2},
\ee
i.e. a subdiffusive behavior, which signifies that (for the time scales
considered here) the trajectories of the chain's beads are spatially more
confined that there of simple Brownian particle.

Now, one can readily find that the optimal trajectory $\tilde{X}_{\rm
sl}(\tau)$ which minimizes the action in Eq.(\ref{u}) obeys the following
Euler equation
\be \label{min_ac1}
\frac{d}{d \tau}\left(\left(\frac{\beta K}{D}\right)^{1/2}
\int_0^{\tau}\dot{\tilde{X}}_{\rm sl}(\tau')\frac{d \tau'}
{\sqrt{\tau-\tau'}}+\lambda\right)=0
\ee
We seek the solution of Eq.(\ref{min_ac1}) in the form
$$
\dot{\tilde{X}}_{\rm sl}(\tau')=A \; (\tau')^{\alpha}
$$
where $\tau'=\tau u$ and $0\le u\le 1$. Substituting this form into
Eq.(\ref{min_ac1}), we get the functional equation
\be \label{anz}
\left(\frac{\beta K}{D}\right)^{1/2}\;
A\;\tau^{\alpha+\frac{1}{2}}\int_0^1 \frac{u^{\alpha}\;du}{\sqrt{1-u}}+
\lambda=C
\ee
where $C$ is some constant. Note now, that the left--hand side of
Eq.(\ref{anz}) is independent of $\tau$ for $\alpha=-1/2$ only, which thus
fixes the value of $\alpha$ to $\alpha = - 1/2$. Hence, we obtain
\be \label{vel}
\dot{\tilde{X}}_{\rm sl}(\tau)=\frac{C-\lambda}{\pi}\;
\left(\frac{D}{\beta K \tau}\right)^{1/2}
\ee
which leads to the following expression for the minimal action
\be \label{ac_min}
S\Big\{\tilde{X}_{\rm sl}(\tau)\Big\}=\frac{2}{\pi}
\left(\frac{D t}{\beta K}\right)^{1/2} (C-\lambda)^2+
\lambda \left(x_{0}-x_{0}(0)\right)
\ee
In turn, the value of the constant $C$ can be found by integrating
Eq.(\ref{vel}), which gives:
$$
C= \lambda + \frac{\pi}{2} \left(\frac{\beta K}{D t}\right)^{1/2}
\Big(x_{0}-x_{0}(0)\Big)
$$
Substituting this expression into Eq.(\ref{ac_min}), we arrive at the final
equation for the minimal action
\be \label{ac_min1}
S\Big\{\tilde{X}_{\rm sl}(\tau)\Big\}= \frac{\pi}{2} \left(\frac{\beta K}{D
t}\right)^{1/2} \Big(x_{0}-x_{0}(0)\Big)^2 + \lambda
\left(x_{0}-x_{0}(0)\right)
\ee

Next, taking advantage of Eqs.(\ref{ac_min1}) and (\ref{1d_path_pol}), we find
that the probability that the slip--link which is at point $x_0$ at $t = 0$
will stay inside the interval $[0,L]$ until time $t$ is given to
exponential accuracy by
\be \label{mirror}
\Psi_{\rm sl}\Big(x_0,x_0(0),t|L\Big) \approx
\sum_{m=-\infty}^{\infty} \exp\left\{- \frac{2}{\pi} \left(\frac{\beta K}{D
t}\right)^{1/2}
\Big(x_0-x_0(0)-2mL\Big)^2\right\},
\ee
which yields, by virtue of the "mirror principle", the following result
\be \label{a}
\Psi_{\rm sl}\Big(x_0,x_0(0),t|L\Big) \approx
\sum_{m=-\infty}^{\infty}
\exp\left\{-\frac{\pi m^2}{2 L^2} \left(\frac{D t}{\beta
K}\right)^{1/2}\right) \sin\left(\frac{\pi m
x_0}{L}\right)\sin\left(\frac{\pi m x_0(0)}{L}\right\}
\ee
Further on, integrating Eq.(\ref{u}) over $x_0$ and $x_0(0)$, we get for the
position-averaged function $\Psi_{\rm sl}(t|L)$ (see Eq.(13))
\be \label{polytrap}
\Psi_{\rm sl}(t|L) \approx \sum_{l=0}^{\infty}(2l+1)^{-2}
\exp\left\{-\frac{\pi(2l+1)^2}{2L^2}
\left(\frac{D t}{\beta K}\right)^{1/2}\right\},
\ee
and consequently, the desired probability $P_{\rm sl}(t)$ that the slip--link
will not encounter any of the traps until time $t$ is given by the following
integral
\be \label{surv_new}
P_{\rm sl}(t) \approx \sum_{l=0}^{\infty}(2l+1)^{-2}
\int_0^{\infty} dL\; \exp\left(-\frac{\pi^2 (2l+1)^2 }{L^2} D
\theta - n_{\rm tr}L\right)
\ee
where for notational convenience we have introduced the "effective" time
$\theta$, where $\theta = (t/\beta K D)^{1/2}/2\pi$. Note that
Eq.(\ref{surv_new}) becomes identical to Eq.(\ref{survival}) upon the mere
replacement $t \to \theta$, which readily enables us to get the corresponding
decay forms from Eqs.(\ref{inte}). We recall, however, that Eq.(\ref{surv_new})
is valid only for $t < \tau_{R}$, which will result in a slightly more
complicated overall decay pattern than the one described by Eqs.(\ref{inte}).

Consider next the evolution of $\Psi_{\rm sl}\Big(x_0,x_0(0),t|L\Big)$ and,
respectively, of $P_{\rm sl}(t)$ in the limit $t > \tau_{R}$. Note that here
the action in Eq.(\ref{newaction}) reduces to the standard result
\be \label{ac_lim}
S\Big\{\tilde{X}_{\rm sl}(\tau)\Big\}=\frac{N}{4 D}\int_0^t d\tau
\left(\frac{d \tilde{X}_{\rm sl}(\tau) }{d \tau}\right)^2 + \lambda
\int_0^t d\tau \left(\frac{d \tilde{X}_{\rm sl}(\tau)}{d \tau}\right),
\ee
which is simply the action of an isolated Brownian particle which moves
with the diffusion coefficient $D/N$ (compare to Eq.(\ref{1d_act})).
Consequently, in this time limit we have that the mean-square displacement
of the slip--link obeys
\be \label{displong}
\overline{x_{0}^2(t)} \sim \frac{D}{N}\,t
\ee
and the  probability $P_{\rm sl}(t)$ follows
\be \label{in}
P_{\rm sl}(t) \; \approx \; \sum_{l=0}^{\infty}(2l+1)^{-2}
\int_0^{\infty} dL\; \exp\left(-\frac{\pi^2 (2l+1)^2 D t}{N
L^2}-n_{\rm tr}L\right)
\ee
Note that this result could be expected on intuitive grounds, since, as we have
already mentioned, in the limit $t > \tau_{R}$ the motion of every bead of the
chain follows mainly that of chain's center--of--mass.

\subsection{Trapping pattern $P_{\rm sl}(t)$ for the slip--link of a Rouse
chain.}

Consider first the case of an infinitely long chain, then $\tau_{R} = \infty$
and the dynamics of the slip--link is described by
Eqs.(\ref{triu2}),(\ref{i}.B) and (\ref{disp}) over the entire time domain.
Comparing the decay forms in Eqs.(\ref{survival}),(\ref{inte}) and
(\ref{surv_new}), we readily find that the probability that the slip--link
will not encounter any of the traps until time $t$ shows the following two
stage decay pattern:
\be \label{slinte}
P(t) \; \approx \; \left\{\begin{array}{lll} \exp\left(-\frac{2}{\pi}
n_{\rm tr} \Big(4 D t/\beta K\Big)^{1/4}\right), & \mbox{for $t_{m} \ll t
\ll \tilde{t}_{\rm c,1}$}, & \quad (A) \\ \exp\left(-\frac{3}{2} n_{\rm
tr}^{2/3} \Big(\pi^2 D t/\beta K \Big)^{1/6}\right), & \mbox{for $t \gg
\tilde{t}_{\rm c,1}$}, & \quad (B) \end{array}\right.
\ee
In Eq.(\ref{slinte}) the crossover time $\tilde{t}_{\rm c,1}$ separating
the regimes A and B obeys $\tilde{t}_{\rm c,1} \approx \beta K/D n_{\rm
tr}^4$. Note that $\tilde{t}_{\rm c,1} \approx (b n_{tr})^{-2} t_{c}$,
where $b$ is the mean distance between the chain's beads in equilibrium and
$t_{c}$ is the corresponding crossover time in Eq.(\ref{inte}). Hence, for
systems with a small density of traps $\tilde{t}_{\rm c,1}$ can be
significantly larger than the monomer crossover time $t_{c}$ in
Eq.(\ref{inte}). Note also that the Eq.(\ref{slinte}.A) is the mean--field,
Smoluchowski--type result corresponding to the sub-diffusive motion of the
slip--link, described by Eq.(\ref{disp}), in presence of uniformly
distributed traps; the exponent in Eq.(\ref{slinte}.A) is just the product
of the trap mean density and the mean maximal range (span) of the
slip--link displacement. On the other hand, Eq.(\ref{slinte}.B) stems from
the interplay between exponentially rare, large trap--free voids, (the
large--$L$ tail of Eq.(\ref{poiss})), and anomalously confined trajectories
of the slip--link. Equation (\ref{slinte}.B) also describes the long--time
tail of the moment generating function of the Wiener sausage volume for the
the slip--link trajectories of an infinitely long chain.

We turn next to the case of finite chains, and hence to finite $\tau_{R}$,
which sets the upper bound on the time of applicability of the decay
pattern in Eq.(\ref{slinte}). Here, at times greater than $\tau_{R}$ the
conventional diffusive motion of the slip--link is restored,
Eqs.(\ref{triu2}),(\ref{i}.A) and (\ref{displong}), and the decay has a
form similar to that in Eqs.(1),
\be \label{sli}
P(t) \; \approx \;
\left\{\begin{array}{lll}
\exp\left(-4 n_{\rm tr} (D t/\pi N)^{1/2}\right), &
\tau_{R} \ll t \ll \tilde{t}_{\rm c,2}, & \quad (A) \\
\exp\left(-3 (\pi^2 n_{\rm tr}^2 D t/4 N)^{1/3}\right), &
t \gg \tilde{t}_{\rm c,2}. & \quad (B) \end{array}\right.
\ee
Here the crossover time $\tilde{t}_{\rm c,2}$ between the A and B regimes is
given by $\tilde{t}_{\rm c,2} \approx N/D n_{tr}^2 = N t_{c}$.

Note, however, that the overall decay pattern of $P_{\rm sl}(t)$ is not the
does not necessarily follow sequentially after Eq.(\ref{slinte}), i.e. the
sequence given by Eqs.(\ref{slinte}.A), (\ref{slinte}.B), (\ref{sli}.A) and
finally, (\ref{sli}.B) may be realized only if the crossover times would obey
the following  multiple inequality $t_{m} \ll \tilde{t}_{c,1} \ll \tau_{R} \ll
\tilde{t}_{c,2}$ which practically is never the case. To show this explicitly
and to construct the actual overall decay pattern, it is expedient to rewrite
the crossover times $\tilde{t}_{c,1}$ and $\tilde{t}_{c,2}$ in terms of the
Rouse time $\tau_{R}$. We have then $\tilde{t}_{c,1} \approx Q^{-4} \tau_{R}$
and $\tilde{t}_{c,2} \approx Q^{-2} \tau_{R}$, where $Q = n_{\rm tr} (b
N^{1/2})$, i.e. is equal to the mean number of traps in the area covered by
a Rouse chain of arclength $b N$ in its typical equilibrium configuration.

Below we analyse different possible situations with respect to the values
of the parameters $Q$ and $q = n_{\rm tr} b$ and discuss the corresponding
decay patterns.
\medskip

{\rm \sc Case I. High density of traps, $q \sim 1$, and long chains, $Q \ll 1$.}
Note first that in this case the crossover time $\tilde{t}_{c,1}$ is comparable
to the microscopic diffusion time, i.e. $\tilde{t}_{c,1} \sim t_{m} \sim b^2/D$,
which implies that the Smoluchowski--type regime corresponding to the
sub-diffusive slip--link motion, Eq.(\ref{slinte}.A), is unobservable. Hence,
the law in Eq.(\ref{slinte}.B) will describe the decay of $P_{\rm sl}(t)$
until $t \approx \tau_{R}$. Further on, the condition $Q \gg 1$ implies
that $\tilde{t}_{c,2} \ll \tau_{R}$ and the regime described by
Eq.(\ref{sli}.A) does not exist. Consequently, the overall decay pattern in
the case I reads
\be \label{case1}
P_{\rm sl}(t) \; \approx \;
\left\{\begin{array}{lll}
\exp\left(-\frac{3}{2} n_{\rm tr}^{2/3}
\Big(\pi^2 D t/\beta K \Big)^{1/6}\right),
& t_{m} \ll t \ll \tau_{R}, & \quad (A) \\
\exp\left(-3 n_{\rm tr}^{2/3} (\pi^2  D t/4 N)^{1/3}\right), &
t \gg \tau_{R}, & \quad (B) \end{array}\right.
\ee
Note that both A and B regimes are essentially non-mean--field and stem
from the presence of fluctuation trap--free voids. We also remark that in
Eqs.(\ref{case1}) the most representative regime is the one in
Eq.(\ref{case1}.A), which is associated with the sub-diffusive behavior of the
slip--link, Eq.(\ref{disp}). Note that the value of $P_{\rm sl}(t)$ at the
crossover time separating the A and B regimes, i.e. $P_{\rm sl}(t=
\tau_{R})$, is of order of $\exp(-Q^{2/3}) \ll 1$, which means that the
probability that the slip--link will be trapped during the stage described
by Eq.(\ref{case1}.A) is considerably higher than the probability that it
will be trapped according to the law in Eq.(\ref{case1}.B).
\medskip

{\rm \sc Case II. Low density of traps, $q \ll 1$, and very long chains,
$Q \gg 1$.} In this case $\tilde{t}_{c,1} \gg t_{m}$, which implies that the
regime in Eq.(\ref{slinte}.A) describes the initial kinetic stage. Next,
since here $\tilde{t}_{c,1} \ll \tau_{R}$, the regime in Eq.(\ref{slinte}.B)
will also exist and will describe the intermediate--time kinetic behavior 
of $P_{\rm sl}(t)$. Lastly, the final stage will follow the decay in
Eq.(\ref{sli}.B), because $\tilde{t}_{c,2}$ appears to be much less than
the Rouse time $\tau_{R}$ and thus the regime in Eq.(\ref{sli}.A) will be
absent. Consequently, in case II one has the following overall decay
pattern:

\be \label{case2}
P_{\rm sl}(t) \; \approx \; \left\{\begin{array}{llll}
\exp\left(-\frac{2}{\pi} n_{\rm tr} \Big(4 D t/\beta K\Big)^{1/4}\right),
& t_{m} \ll t \ll \tilde{t}_{\rm c,1}, & \quad (A) \\
\exp\left( - \frac{3}{2} n_{\rm tr}^{2/3}
\Big(\pi^2 D t/\beta K \Big)^{1/6}\right), &
\tilde{t}_{\rm c,1} \ll t \ll \tau_{R}, & \quad (B) \\
\exp\left(-3 (\pi^2 n_{\rm tr}^2 Dt/4 N)^{1/3}\right), &
t \gg \tau_{R}, & \quad (C) \end{array}\right.
\ee
In Eqs.(\ref{case2}) the decay laws in the first two lines are associated with
the sub-diffusive motion of the slip--link and describe the mean-field,
Smoluchowski-type (A) and fluctuation--induced (B) kinetic stages, 
respectively, while the law in the third line describes the survival of a 
Brownian particle (with diffusion coefficient $D/N$) in the fluctuation 
trap--free voids. Note that here the most representative regime is the one 
associated with the sub-diffusive motion in the slip--link and fluctuation 
trap--free voids, namely, regime (B). As in case I, we have here $P_{\rm 
sl}(t = \tau_{R}) \approx \exp(-Q^{2/3})$, which is very small. This 
implies that the most probable decay is given by Eq.(\ref{case2}.B). The 
first stage, i.e. the decay described by  Eq.(\ref{case2}.A), appears to be 
relatively unimportant, since during this stage $P_{\rm sl}(t)$ does not 
drop appreciably, $P_{\rm sl}(t = \tilde{t}_{\rm c,1}) \approx 1/3$.
\medskip

{\rm \sc Case III. Low density of traps, $q \ll 1$, and short chains,
$Q \ll 1$.} In this particular situation we have that $\tilde{t}_{\rm c,1}
\gg t_{m}$ and consequently, the initial decay obeys Eq.(\ref{slinte}.A).
Further on, since here $\tilde{t}_{\rm c,1}$ also exceeds the Rouse time, i.e.
$\tilde{t}_{\rm c,1} \gg \tau_{R}$, the regime predicted by Eq.(\ref{slinte}.B)
is absent, which means that the decay  in Eq.(\ref{slinte}.A) crosses over at
$t = \tau_{R}$ to the decay predicted by  Eq.(\ref{sli}.A). Lastly, the
stretched--exponential dependence in Eq.(\ref{sli}.A) is followed at
$t > \tilde{t}_{\rm c,2}$, $\tilde{t}_{\rm c,2} \gg \tau_{R}$ by the form
Eq.(\ref{sli}.B). Hence, in case III one has that $P_{\rm sl}(t)$ follows
\be \label{case3}
P_{\rm sl}(t) \; \approx \; \left\{\begin{array}{llll}
\exp\left(-\frac{2}{\pi} n_{\rm tr} \Big(4 D t/\beta K\Big)^{1/4}\right),
& t_{m} \ll t \ll \tau_{R}, & \quad (A) \\
\exp\left(-4 n_{\rm tr} (D t/\pi N)^{1/2}\right), &
\tau_{R} \ll t \ll \tilde{t}_{\rm c,2}, & \quad (B) \\
\exp\left(-3 n_{\rm tr}^{2/3} (\pi^2 D t/4 N)^{1/3}\right), &
t \gg \tilde{t}_{c,2}, & \quad (C) \end{array}\right.
\ee
In this case, however, only the last, fluctuation--induced regime C appears to
be significant; one can readily verify that $P_{\rm sl}(t)$ practically 
does not change during the regimes A and B, $P_{\rm sl}(t=\tilde{t}_{\rm 
c,2}) \approx \exp(-4/\pi) \sim 1$.

\section{Conclusions}

To summarize, we have studied the dynamics of an isolated Rouse chain, which
diffuses in a three--dimensional space under the constraint that one of its
extremities, the slip--link, may move only along a line containing randomly
placed immobile traps. For such a model we have computed exactly the time
evolution of the probability $P_{\rm sl}(t)$ that the slip--link will not
encounter any of the traps until time $t$, i.e. that the chain will remain
completely mobile until this moment of time. We have shown that in the most
general case this probability is a succession of several
stretched--exponential functions of time, where the dynamical exponents
depend on the time of observation and on characteristic crossover times. We
have specified these crossover times and have determined explicitly the
forms of $P_{\rm sl}(t)$ in several particular situations.
\bigskip

\noindent{\bf Acknowledments}
\medskip

We are very thankful to Prof. M. Moreau for discussions and acknowledge the
support of the PROCOPE--DAAD programm of the DFG through SFB--428 and of the 
Fonds der Chemischen Industrie.

\end{document}